\documentclass[12pt]{article}
\usepackage{amsmath}
\usepackage{amssymb}
\usepackage{amscd}
\usepackage{amsfonts}
\usepackage{amsthm}
\usepackage{authblk}

\usepackage{mathrsfs}
\usepackage{amsbsy}
\usepackage{amssymb}
\makeindex
  \newtheorem{theorem}{Theorem}[section]
  \newtheorem{prop}[theorem]{Proposition}
  \newtheorem{lemma}[theorem]{Lemma}
  \newtheorem{cor}[theorem]{Corollary}
\theoremstyle{remark}
  
  \newtheorem{rem}[theorem]{Remark}

\def\bea{\begin{eqnarray}}
\def\eea{\end{eqnarray}}

 \newcounter{biscompt}

 \theoremstyle{plain}

\usepackage[hidelinks]{hyperref}

 \numberwithin{equation}{section}

\newcommand{\Dj}{\hbox to 8pt{\raisebox{.4\height}{-}\hss D}}

\newcommand{\beq}[1]{\begin{equation}\label{#1}}
\newcommand{\eq}{\end{equation}}
\newcommand\eeq{\end{equation}}
\newcommand\bqa {\begin{eqnarray}}
\newcommand\eqa {\end{eqnarray}}

\newcommand\nn {\nonumber}

\newcommand{\bear}{\begin{array}}
\newcommand{\enar}{\end{array}}

\newcommand{\R}{\mathbb{R}}

\usepackage[usenames]{color}

\usepackage{tikz}
\usepackage{tikz-cd}

\usepackage{graphicx}

\title{Lie-Bianchi integration of the full symmetric Toda system and the Lanczos transform}
\author[1,2,3]{Yury B. Chernyakov\footnote{chernyakov@itep.ru}}
\author[1,2,4]{Georgy I. Sharygin\footnote{sharygin@itep.ru}}
\author[1,4,5]{Dmitry V. Talalaev\footnote{dtalalaev@yandex.ru}}
\affil[1]{\small NRC ``Kurchatov institute'', Kurchatov square, 1, 123182, Moscow, Russia}
\affil[2]{\small  MIPT, Institutskii per. 9, 141700, Dolgoprudny, Russia}
\affil[3]{\small 
IITP RAS, Bolshoy Karetny per. 19, build.1, 127051, Moscow, Russia}
\affil[4]{\small  Lomonosov MSU, Leninskie Gory, 1, 119991, Moscow, Russia}
\affil[5]{\small Demidov YarSU, Sovetskaya Str. 14, 150003, Yaroslavl, Russia}
\date{}

\begin{document}

\maketitle

\begin{abstract}
In our previous work \cite{CST26} we described the action of the stochastic Lie algebra by vector fields on the group $SO_d(\R)$, which helped to prove the Lie-Bianchi criterion for the full symmetric Toda system. In this paper we continue the discussion by constructing a related action of the stochastic group $ST_d(\R)$ on $SO_d(\R)$; we further use this action to get an explicit dynamics-preserving map from the generic isospectral locus of the full symmetric Toda system (consisting of real symmetric $d\times d$ matrices) to the space of symmetric tridiagonal matrices, also called Jacobi matrices. The construction is based on the systematic use of the Lanczos--Krylov map. We also give a reconstruction procedure that avoids extracting the individual eigenvalues and eigenvectors.
\end{abstract}

\tableofcontents

\section{Introduction}
The full symmetric Toda system is an important generalisation of the usual open Toda chain; it is a system of differential equations usually determined in terms of its Lax representation:
\begin{equation}\label{eq:Toda1}
\dot L=[L,M(L)],\qquad L\in\mathrm{Sym}_d(\R),\qquad M(A)=A_- - A_-^T,
\end{equation}
where $A_-$ denotes the strictly lower-triangular part of an arbitrary matrix $A$. Thus $M$ is the projection of $\mathfrak{gl}_d(\R)$ onto $\mathfrak{so}_d(\R)$ along the \emph{upper}-triangular subalgebra. Observe that in most other sources the right hand side of the equation has opposite sign; we use this sign convention here for the sake of convenience and compatibility with our further calculations, an interested reader can make the necessary corrections.
Evidently the equation preserves symmetry of matrices and their spectrum, and on symmetric tridiagonal matrices restricts to the usual finite nonperiodic, or open, Toda chain. Some information about these systems can be found in the review \cite{Reviews}; we also refer the reader to the foundational works \cite{Flaschka1974,Moser1975,Ar1979,K1978,S1980}, and \cite{DLNT} for the discussion of inverse-spectral and generic-orbit approach .

The full symmetric system has been extensively studied by many authors in the course of last 50 years, see \cite{Reviews}. It's been shown that it admits Hamiltonian description: the Poisson structure on the space of symmetric matrices is induced from an isomorphism with the (dual space of) upper triangular matrices, and the Hamilton function is taken equal to the trace of $L^2$. On generic symplectic leaves this system is Liouville integrable: the first integrals given by invariant functions on the space of matrices are not sufficient for integrability, and additional first integrals are rational functions of the matrix coefficients, see \cite{DLNT,GS1999,CS14,CS15}. This system is closely related with the geometry of flag manifolds, Bruhat and Schubert stratifications: in particular its trajectories on isospectral manifolds (the spectrum of the matrix is invariant of the system) preserve the Schubert cells and follow the Bruhat order. The same holds for its generalizations to other root systems/ real semisimple Lie groups. Its solutions are related to QR-decompositions and to the factorisation methods described in \cite{S1980,TCS23}.

In addition to being Liouville-integrable, this system has many other integrability properties: it is written in Lax form (however, without spectral parameter), it is super-integrable in the sense of Nekhoroshev (i.e. it has a sufficiently large non-involutive family of first integrals, see \cite{CS14}) and it is also integrable in the sense of Lie-Bianchi test, i.e. it admits a solvable Lie algebra of infinitesimal symmetries satisfying the required local rank condition. The latter fact has been the topic of our previous works, \cite{CST26,CSS23}, where we described a finite-dimensional Lie algebra of vector fields commuting with Toda system (or rather with its pull-back on the group of orthogonal matrices). One can now choose in this algebra a solvable subalgebra of sufficiently large dimension: this can either be done ``by hand'' (it is possible to pick the corresponding vector fields explicitly), or by observing that the algebra in question is equal to a central extension of the so-called ``stochastic Lie algebra'', i.e. the Lie algebra of square matrices $A$ such that $Ae=0$, where $e=(1,\dots,1)^T$. This Lie algebra is embedded into $\mathfrak{gl}_d$, and hence its intersections with various Borel subalgebras can be used for our purposes. It is essential, however, to verify the independence of the selected fields on the relevant invariant manifold; solvability alone does not imply the Lie--Bianchi criterion \cite{Lie1891,Carinena2015,Carinena2016}.

In this paper we continue the study started in \cite{CST26,CSS23}. We begin with recalling the main constructions from our previous works: we reformulate the Toda equation as the dynamical system induced by a vector field on orthogonal group; then we recall the construction of infinitesimal symmetries of the system from \cite{CSS23}. We use a realisation adapted to the normalised RQ-decomposition and single out a finite-dimensional Lie algebra of symmetries associated with stochastic matrices (we recall their definition below). In Section \ref{St_gr_act} we describe the corresponding \textit{stochastic group action} on the space of orthogonal matrices. This is the first important result of our paper. With the projection $M$ fixed in \eqref{eq:Toda1}, its infinitesimal generators agree with the stochastic fields defined below. The relation with the different projection conventions used in earlier accounts should be kept distinct from this internal identification; compatibility with the Toda flow will also be established by factorisation. In the next section we recall the \textit{Lanczos-Krylov map}, that sends a symmetric matrix with a cyclic initial vector to a three-diagonal symmetric matrix (such matrices are often called \textit{Jacobi matrices}). We show that if the initial vector of this procedure is taken to be $e_d$ (the last basis vector in $\R^d$) then this transformation can be realised via the action of stochastic group that we described above. In particular, this means that \textit{the Lanczos--Krylov map intertwines the full symmetric Toda flow and the open Toda flow on the cyclic simple-spectrum locus}; the original trajectory can subsequently be restored from the tridiagonal solution. In the last section we describe the inverse procedure, that allows one to restore the solution of the full Toda system from the solution of the open Toda chain. At the end we make several speculative remarks concerning possible generalisations and corollaries of our construction.

\section{Basic constructions and preliminaries}
In this section we remind the main constructions concerned with the full symmetric Toda system, as well as some of our previous results. First of all, recall that one can reinterpret the full symmetric Toda system \eqref{eq:Toda1} as a dynamic system on the orthogonal group.  Namely let $L_0$ be a real symmetric matrix, and let $\Lambda=\Lambda_0$ be the diagonal matrix of eigenvalues of $L_0$: then
\[
L_0=\Psi_0\Lambda\Psi_0^T
\]
for some orthogonal matrix $\Psi_0$. The matrix $\Psi_0$ is uniquely determined, up to the action of a discrete group, so it is locally unique and without loss of generality we may assume that $\det\Psi_0=1$. Consider the following vector field on $SO_d(\mathbb R)$:
\[
\mathcal X_\Lambda(\Psi)=-M(\Psi\Lambda\Psi^T)\Psi.
\]
Here and throughout the paper $M(A)=A_--A_-^T$ is the \emph{same}
projection, onto skew-symmetric matrices along the upper-triangular
subalgebra, as in \eqref{eq:Toda1}. For a symmetric matrix
$L=\Psi\Lambda\Psi^T$ we have $M(L)=L_--L_+$.
It is now easy to see that the following is true:
\textit{if $\Psi(t)$ is the trajectory of $\mathcal X_\Lambda$
with $\Psi(0)=\Psi_0$, then the matrix
$L(t)=\Psi(t)\Lambda\Psi^T(t)$ solves the full symmetric Toda
system with initial condition $L(0)=L_0$.}

Now we can speak about the Toda system in terms of the properties of
$\mathcal X_\Lambda$: instead of looking for symmetries and first
integrals of the original equation, one can look for the
invariants and symmetries of this field. The works
\cite{CST26,CSS23} supply the general background on infinitesimal
symmetries and Lie--Bianchi integrability. For the sake of completeness, let us briefly recall the results from these papers here.
\begin{theorem}
\label{thm:RQfields}
\begin{enumerate}
\item[{\em(\textit{i})}] For any $X\in\mathfrak{gl}_d(\R)$ put
\[
\mathcal T^X(\Psi)=M(\Psi X\Psi^T)\Psi.
\]
The map $X\mapsto\mathcal T^X$ defines a Lie algebra
representation by vector fields:
\[
[\mathcal T^X,\mathcal T^Y]=\mathcal T^{[X,Y]},
\qquad \mathcal T^I=0.
\]
The lifted Toda field is $\mathcal X_\Lambda=-\mathcal T^\Lambda$.
\item[{\em(\textit{ii})}] On the open set where every entry
$w_i=\psi_{di}$ is nonzero, the fields
\[
T_{ij}(\Psi)=\frac{w_j}{w_i}\mathcal T^{E_{ij}}(\Psi),
\qquad i,j=1,\ldots,d,
\]
commute with the lifted Toda field $\mathcal X_\Lambda$.
\item[{\em(\textit{iii})}] The linear span of the $T_{ij}$ is
closed under Lie brackets. In particular, for $i\ne j$ the fields
\[
\xi_{ij}=T_{ij}-T_{jj}
\]
are the infinitesimal generators of the left-stochastic action
defined in Section~\ref{St_gr_act}.
\end{enumerate}
\end{theorem}
Observe that here we use the convention adapted with the RQ-decompositon (\textit{not} QR-decomposition), which is directly compatible with the action constructed in the next section.
\begin{proof}
For the sake of completeness, let us briefly recall the proof of these results. The normalised RQ-factorisation of $\Psi\exp(tX)$ gives a right
$GL_d^+(\R)$-action on $SO_d(\R)$; its infinitesimal generator is
$\mathcal T^X$, which proves (i). For (ii), the last row of
$\mathcal X_\Lambda$ satisfies
\begin{equation}\label{eq:weightdiff}
\dot w_i=(L_{dd}-\lambda_i)w_i,
\qquad
\mathcal X_\Lambda\!\left(\frac{w_j}{w_i}\right)
=(\lambda_i-\lambda_j)\frac{w_j}{w_i}.
\end{equation}
Since
$[\mathcal X_\Lambda,\mathcal T^{E_{ij}}]
=(\lambda_j-\lambda_i)\mathcal T^{E_{ij}}$,
the two contributions to
$[\mathcal X_\Lambda,T_{ij}]$ cancel.
The last assertion follows from the identification of $\xi_{ij}$
with the left-stochastic infinitesimal generators below, together
with the fact that the latter action commutes with the positive
diagonal RQ-action; see the proof after
\eqref{eq:infinitesimal-new}.
\end{proof}

The Lie algebra of these fields is a natural stochastic symmetry
algebra for the RQ-adapted convention that we use in this paper. Indeed, its generators are
built from a representation of $\mathfrak{gl}_d$, while the
$\xi_{ij}$ correspond to the elementary left-stochastic matrices
$E_{ij}-E_{jj}$. The diagonal fields $T_{jj}=\mathcal T^{E_{jj}}$
form an abelian algebra of dimension $d-1$. The combination of
these infinitesimal symmetries with solvable stochastic
subalgebras was earlier used in the papers
\cite{CST26,CSS23}; he fact that the rank of the corresponding solvable Lie algebra is sufficient is discussed in \cite{CST26}: 
the solvability of the abstract algebra alone
is insufficient \cite{Lie1891,Carinena2015,Carinena2016}.

\section{The stochastic group action}

\subsection[The rational action on GL(d,R)]{The rational action on $GL_d(\R)$}
\label{St_gr_act}
Recall that a matrix $A\in\mathrm{Mat}_d(\R)$ is called left- (resp. right-) \textit{stochastic}, if the sum of the elements in each column (resp. in each row) of $A$ is equal to $1$. The term ``stochastic'' should not mislead the reader: we are not assuming positivity of matrix elements only their sums are fixed. It is not hard to see that if both $A$ and $B$ are left- (resp. right-) stochastic, then so is their product: in effect the definition of left-/right- stochastic matrix $A$ is equivalent to the condition that the left (resp. right) multiplication of $A$ by the row (resp. column) vector $(1,1,\dots,1)$ (resp. $(1,1,\dots,1)^T$) preserves this vector. A matrix is called \textit{bi-stochastic} if it's both left and right stochastic.

It is clear from the definitions that the set of invertible left-, right- or bi-stochastic matrices form a subgroup in the group $GL_d(\R)$. We will refer to these groups as left-, right- or bi-stochastic groups respectively or just \textit{stochastic groups} unless it can cause confusion and denote them by $ST^\ell_d(\R),\,ST^r_d(\R),\,ST^{bi}_d(\R)$ respectively. If no confusion is caused, we will also denote the groups $ST^\ell_d(\R),\,ST^r_d(\R)$ simply by $ST_d(\R)$; this notation is justified by the observation that the map $g\mapsto(g^T)^{-1}$ for any $g\in GL_d(\R)$ induces an isomorphism $ST^\ell_d(\R)\to ST^r_d(\R)$. The Lie algebras of these groups are equal to the spaces:
\[
\begin{split}
\mathfrak{st}_d^\ell&=\Bigl\{X=(x_{ij})\in\mathrm{Mat}_d(\R)\Big| \sum_{i=1}^dx_{ij}=0,\,j=1,\dots,d\Bigr\},\\
\mathfrak{st}_d^r&=\Bigl\{X=(x_{ij})\in\mathrm{Mat}_d(\R)\Big| \sum_{j=1}^dx_{ij}=0,\,i=1,\dots,d\Bigr\}
\end{split}
\]
and $\mathfrak{st}_d^{bi}=\mathfrak{st}_d^\ell\bigcap \mathfrak{st}_d^r$. It is easy to see that the map $X\mapsto -X^T$ induces an isomorphism $\mathfrak{st}_d^\ell\cong\mathfrak{st}_d^r$; as in the case of groups, we will usually denote both these algebras simply by $\mathfrak{st}_d$.

The group $ST_d(\R)$ is a subgroup in $GL_d(\R)$, and hence it acts on $GL_d(\R)$ by multiplications both from the left and from the right. We are now going to modify this action a little, and obtain an action of the stochastic group on a subset of $GL_d(\R)$; we can also regard this construction as a \textit{rational} action of $ST_d(\R)$ on $GL_d(\R)$, i.e. each $g\in ST_d(\R)$ will induce \textit{rational} (in the sense of algebraic geometry) transformation on $GL_d(\R)$; recall that \textit{rational morphisms} are determined by rational transformations of coordinates, in particular they can have poles, where denominators of the corresponding rational functions vanish. In our case we will use the canonical global coordinates on $GL_d(\R)$ induced from its inclusion into the Euclidean space of matrices.

So let $x_1,\dots,x_d\in\R$ be a $d$-tuple of real numbers; we assume that $x_i\neq 0,\,i=1,\dots,d$. Consider the column vector $v=(x_1,\dots,x_d)^T\in\R^d$. Clearly
\[
v=\begin{pmatrix}x_1 &   0   & \dots & 0\\
	0    & x_2 & \dots & 0\\
	\vdots&    &\ddots& \vdots\\
	0    &  \dots &\dots & x_d
\end{pmatrix}\begin{pmatrix} 1\\ 1\\ \vdots\\ 1\end{pmatrix}.
\]
Denoting the diagonal matrix in this formula by $\Lambda_v$ we see that for any right-stochastic matrix $A$ the following equation holds
\[
v=\Lambda_v A\Lambda_v^{-1}v.
\]
Now for any invertible diagonal matrix $D$ we have
\[
D v=\Lambda_{D v}A\Lambda_{D v}^{-1}D v.
\]
Dually, if $w=(y_1,\dots,y_d)$ is a row vector such that $y_i\neq0,\,i=1,\dots,d$, then for every left-stochastic matrix $B$ and every invertible diagonal matrix $D$ we have in a similar notation
\[
\begin{split}
	w&=w\Lambda_w^{-1}B\Lambda_w\\
	w D&=(w D)\Lambda^{-1}_{w D} B\Lambda_{w D}.
\end{split}
\]
Now we can define the right rational action of left-stochastic matrices on the space of $d\times d$ matrices: for any $M\in\mathrm{Mat}_d^0(\R)$ and any $B\in ST_d^\ell(\R)$ we put
\begin{equation}\label{eq:actgl}
W\triangleleft B= W\Lambda_{\rho(W)}^{-1}B\Lambda_{\rho(W)},
\end{equation}
where $\rho$ is the projection map
\begin{equation*}
\begin{tikzcd}
	\mathrm{Mat}_d(\R) \arrow{d}{\rho} \\
	\,\,\,\R^{d}
\end{tikzcd}
\end{equation*}
given by taking the bottom row of the matrix, and $\mathrm{Mat}_d^0(\R)$ denotes the space of matrices with non-vanishing entries in the lowest row: the formula \eqref{eq:actgl} is well-defined if the entries in their lower row are nonzero, and in general it can have poles, so for every $B\in ST_d^\ell(\R)$ we get a \textit{rational} transformation of $\mathrm{Mat}_d(\R)$. Moreover, it follows from the definitions and earlier remarks that $\rho(W\triangleleft B)=\rho(W)$, since
\[
\rho(W\triangleleft B)=\rho(W)\Lambda_{\rho(W)}^{-1}B\Lambda_{\rho(W)}=\mathbf{1}^TB\Lambda_{\rho(W)}=\rho(W),
\]
where $\mathbf{1}^T=(1,\dots,1)$, since the matrix $B$ is (left) stochastic. So it follows that
\[
(W\triangleleft B)\triangleleft C=(W\Lambda_{\rho(W)}^{-1}B\Lambda_{\rho(W)})(\Lambda_{\rho(W)}^{-1}C\Lambda_{\rho(W)})=W\triangleleft (BC).
\]
That means that the formula \eqref{eq:actgl} does in fact define an action of the group $ST_d^\ell(\R)$ on $\mathrm{Mat}_d(\R)$ by rational transformations. Moreover, if $W\in\mathrm{Mat}_d^0(\R)$ is invertible, then so is $W\triangleleft B$, and hence we get a \textit{rational action of $ST_d^\ell(\R)$ on the open subset $GL_d(\R)\cap\mathrm{Mat}_d^0(\R)$}. 

For future references we will need the following observation:
\begin{prop}
For any invertible diagonal matrix $D$ we have $(W\triangleleft B)D=(W D)\triangleleft B$; in other words, the action of stochastic group centralises multiplication by diagonal matrices from the right.
\end{prop}
\begin{proof}
We compute:
\[
\begin{split}
	(W\triangleleft B)D &=W\Lambda_{\rho(W)}^{-1}B\Lambda_{\rho(W)}D\\
	&=W D(D^{-1}\Lambda_{\rho(W)}^{-1}B\Lambda_{\rho(W)}D)\\
	&=W D(\Lambda_{\rho(W D)}^{-1}B\Lambda_{\rho(W D)})\\
	&=(W D)\triangleleft B.
\end{split}
\]
\end{proof}

\subsection[The stochastic action on SO(d,R)]{The stochastic action on $SO_d(\R)$}
We are now going to pull the action of $ST_d^\ell(\R)$ from the group
$GL_d(\R)$ to the special orthogonal group $SO_d(\R)$. To this end
we begin with the Gram--Schmidt orthogonalisation process, applied to
the rows in reverse order. It gives the following decomposition:
\begin{equation}\label{eq:decomp1}
GL_d^+(\R)=B_d^+\cdot SO_d(\R),
\end{equation}
where $B_d^+$ is the group of invertible upper triangular matrices with
strictly positive diagonal. This \textit{normalised RQ-decomposition}
$W=RQ$ is unique. In particular the choice of signs, which may be
ambiguous in an unnormalised decomposition, is fixed globally on
$GL_d^+(\R)$.

Let now $\Psi$ be orthogonal and let $g\in GL_d^+(\R)$. We put
\begin{equation}\label{eq:actnew}
\Psi g=R_{\Psi,g}\Psi_g,
\end{equation}
where the right-hand side is the normalised RQ-decomposition.
The uniqueness of this decomposition and the triangularity of the first
factor give
\[
R_{\Psi,gh}\Psi_{gh}
=R_{\Psi,g}R_{\Psi_g,h}(\Psi_g)_h,
\qquad \Psi_{gh}=(\Psi_g)_h.
\]
Thus the right multiplication followed by taking the RQ orthogonal
factor defines an action of $GL_d^+(\R)$ on $SO_d(\R)$.

We are now going to combine this action with the rational action
$\triangleleft$ defined above. From now on $ST_d(\R)$ means the
identity component $ST_d^{\ell,+}(\R)=\{B:\mathbf 1^TB=\mathbf 1^T,
\det B>0\}$. Put
\[
\Omega=\{\Psi\in SO_d(\R):\psi_{di}\ne0\text{ for every }i\}.
\]
For $B\in ST_d(\R)$ and $\Psi\in\Omega$, define
$\Psi\blacktriangleleft B=\Psi_B$ from
\begin{equation}\label{eq:action1}
R_{\Psi,B}\Psi_B=\Psi\triangleleft B
=\Psi\Lambda_{\rho(\Psi)}^{-1}B\Lambda_{\rho(\Psi)},
\quad R_{\Psi,B}\in B_d^+.
\end{equation}
\begin{prop}\label{prop:action}
The map $(\Psi,B)\mapsto\Psi\blacktriangleleft B$ is a globally
defined real-analytic right action of $ST_d(\R)$ on $\Omega$.
Moreover,
\begin{equation}\label{eq:bottom-preserve}
\rho(\Psi\blacktriangleleft B)=\rho(\Psi).
\end{equation}
\end{prop}
\begin{proof}
As observed earlier $\rho(\Psi\triangleleft B)=\rho(\Psi)$.
Since $R_{\Psi,B}$ is upper triangular,
\[
\rho(\Psi\triangleleft B)
=(R_{\Psi,B})_{dd}\rho(\Psi_B).
\]
Both row vectors have norm one and $(R_{\Psi,B})_{dd}>0$;
therefore $(R_{\Psi,B})_{dd}=1$ and $\rho(\Psi_B)=\rho(\Psi)$.
Using this identity, the rational action property and the uniqueness
of normalised RQ gives
\[
(\Psi\blacktriangleleft B)\blacktriangleleft C
=\Psi\blacktriangleleft(BC).
\]
Smoothness and real analyticity follow from the row Gram--Schmidt
algorithm, whose denominators are positive on this domain.
\end{proof}
\begin{rem}
Observe that the action $\triangleleft$ of $ST_d(\R)$ on $GL_d(\R)$ is \textit{rational}; 
however after RQ-orthogonalisation the
$ST_d(\R)$-action on $SO_d(\R)$ is in general only real analytic,
since the normalising factors involve square roots. 
\end{rem}

\subsection{Description of the orbit space}\label{Proj_map}
Let us describe the structure of the orbit space of the action
$\blacktriangleleft$. First, we observe that the action leaves the
bottom row of $\Psi$ fixed. This allows us to use the projection
$\rho:SO_d(\R)\to S^{d-1}$, given by taking the last row. On
$\Omega$ it is a smooth submersion onto
\[
S^{d-1}_{\times}=\{w\in S^{d-1}:w_1\cdots w_d\ne0\}.
\]
The main result of this section is that the fibres of this projection
are 
the orbits of the stochastic group action.
\begin{equation*}
\begin{tikzcd}
\Omega \arrow[r,"\rho"] \arrow[d,two heads] & S^{d-1}_{\times}\\
\Omega/ST_d(\R)\arrow[ur,"\simeq"'] &
\end{tikzcd}
\end{equation*}
\begin{lemma}\label{lemma:orbits}
Two matrices $\Psi,\Phi\in\Omega$ belong to the same orbit if
and only if $\rho(\Psi)=\rho(\Phi)$. In particular,
\[
\dim\mathcal O_{\Psi}=\frac12(d-1)(d-2),\qquad
\Omega/ST_d(\R)\simeq S^{d-1}_{\times}.
\]
\end{lemma}
\begin{proof}
One implication follows from Proposition~\ref{prop:action}.
Conversely, suppose that $\rho(\Psi)=\rho(\Phi)=w$.
Denote $\Lambda_w=\mathrm{diag}(w_1,\ldots,w_d)$ and set
\begin{equation}\label{eq:orbitB}
B=\Lambda_w\Psi^T\Phi\Lambda_w^{-1}.
\end{equation}
Then $\det B=1$ and
\[
\mathbf1^TB=w\Psi^T\Phi\Lambda_w^{-1}
=e_d^T\Phi\Lambda_w^{-1}=\mathbf1^T.
\]
Thus $B\in ST_d(\R)$, while
$\Psi\Lambda_w^{-1}B\Lambda_w=\Phi$.
The normalised RQ-factor of $\Phi$ is $\Phi$ itself;
therefore $\Psi\blacktriangleleft B=\Phi$.
The dimension now follows by subtracting the dimension $d-1$ of the
sphere from $\dim SO_d(\R)=d(d-1)/2$.
\end{proof}
Observe that the proof is constructive: the matrix $B$ is given explicitly by
\eqref{eq:orbitB}, if the matrices $\Psi,\Phi$ are given. 
For future references let us formulate the criterion for two orthogonal matrices to lie in the same stochastic orbit as a separate statement: 
\begin{cor}\label{cor:cor}
For $\Psi,\Psi'\in\Omega$ one has
$\Psi'\in\Psi\blacktriangleleft ST_d(\R)$ if and only if
$\rho(\Psi)=\rho(\Psi')$.
\end{cor}

\subsection{Infinitesimal action}\label{sec:Infinitesimal}
It is our purpose now to identify the vector fields induced by the
stochastic group action. 
Recall that the operator $M$ introduced in the
beginning of the paper is the projection induced by the decomposition:
\begin{equation}\label{eq:projRQ}
\mathfrak{gl}_d(\R)=\mathfrak{b}_d^+\oplus\mathfrak{so}_d(\R),\qquad M:\mathfrak{gl}_d(\R)\to\mathfrak{so}_d(\R),
\end{equation}
so that $M(A)=A_--A_-^T$.

We let $B=B(t)$ be an elementary left-stochastic column
transformation,
$B(t)=I+t(E_{ij}-E_{jj})+O(t^2)$, $i\ne j$.
Differentiating \eqref{eq:action1} at $t=0$ gives
\[
\dot R+\dot\Psi_B\Psi^T
=\Psi\left(\frac{\psi_{dj}}{\psi_{di}}E_{ij}-E_{jj}\right)\Psi^T.
\]
Consequently the infinitesimal field is
\begin{equation}\label{eq:infinitesimal-new}
\begin{aligned}
\xi_{ij}(\Psi)
&=M\!\left(\Psi\left(\frac{\psi_{dj}}{\psi_{di}}E_{ij}
  -E_{jj}\right)\Psi^T\right)\Psi\\
&=T_{ij}(\Psi)-T_{jj}(\Psi).
\end{aligned}
\end{equation}
Thus the stochastic group action \emph{integrates} the RQ-adapted
stochastic infinitesimal fields $\xi_{ij}$. In particular, it preserves the
bottom row pointwise, as can be checked directly: for every
$i,j,k$ one has
\begin{equation}\label{eq:oldvector-on-w}
\mathcal T^{E_{ij}}(w_k)
=w_i\delta_{jk}-w_iw_jw_k,\qquad
T_{ij}(w_k)=w_j\delta_{jk}-w_j^2w_k
\end{equation}
so that 
\[
\xi_{ij}(w_k)=T_{ij}(w_k)-T_{jj}(w_k)=w_j\delta_{jk}-w_j^2w_k-w_j\delta_{jk}-w_j^2w_k=0
\]
for all $k$, in full agreement with \eqref{eq:bottom-preserve}.

\begin{rem}It follows from our reasonings (and also see \cite{CST26,CSS23}), that the (positive) diagonal group action on $SO_d(\R)$, given by the RQ-decomposition of the right product:
\[
R_{\Psi,D}\Gamma_D(\Psi)=\Psi D
\]
has the fields $\mathcal T^{E_{jj}}=T_{jj}$ as infinitesimal generators. It also follows from the earlier remarks that this diagonal action commutes with the stochastic action $\blacktriangleleft$:
\begin{equation}\label{eq:commuting-actions}
\Gamma_D(\Psi\blacktriangleleft B)
=\Gamma_D(\Psi)\blacktriangleleft B
\qquad(D\text{ diagonal with positive entries}).
\end{equation} 
Here positivity is important for the uniqueness of decompositions, otherwise we need to specify the corresponding branch of the maps. By direct computation the equality \eqref{eq:commuting-actions} can be proved as follows: writing $w=\rho(\Psi)$ and
$\Psi D=R_D\Gamma_D(\Psi)$, we have
$\rho(\Gamma_D(\Psi))=wD/\|wD\|$ and
$\Lambda_{wD}=\Lambda_wD$. Using the uniqueness of normalised
RQ-decomposition in the two sides of \eqref{eq:commuting-actions}
then gives the identity.
\end{rem}
It follows that the $\xi_{ij}$ close under Lie brackets as an image of the left-stochastic Lie algebra,
that they commute with every diagonal field
$\mathcal T^{E_{jj}}$, and that the span of all $T_{ij}$ is a
finite-dimensional Lie algebra. This gives an independent proof of
Theorem~\ref{thm:RQfields}(iii).

\begin{rem} 
Since Toda trajectories can be obtained by the RQ-decomposition, the factorisation
argument can also be used to show directly that the stochastic group action preserves these trajectories.
\begin{prop}\label{prop:equivariance}
Let $\Phi_t$ be the lifted Toda flow on $SO_d(\R)$ with fixed ordered eigenvalues $\Lambda$, generated by
$\mathcal X_\Lambda=-\mathcal T^\Lambda$. Then, for any $B\in ST_d(\R)$,
\[
\Phi_t(\Psi\blacktriangleleft B) =\Phi_t(\Psi)\blacktriangleleft B \qquad (\Psi\in\Omega).
\]
In particular, a fixed stochastic group element sends a Toda trajectory to another trajectory with the same spectrum.
\end{prop}
\begin{proof}
It is well-known that for the normalised RQ-factorisation 
\[W=R_W\operatorname{rq}(W),\,R_W\in B_+,\,\operatorname{rq}(W)\in SO_d(\R),
\]
the lifted Toda flow is given by the $SO_d(\R)$ part of the decomposition
\begin{equation}\label{eq:RQevol}
\Phi_t(\Psi)=\operatorname{rq}(\Psi e^{-t\Lambda}).
\end{equation}
With $w=\rho(\Psi)$ we obtain from this equality
\begin{equation}\label{eq:wtformula}
w_i(t)=\frac{w_i e^{-\lambda_i t}}
{\left(\sum_jw_j^2e^{-2\lambda_jt}\right)^{1/2}}.
\end{equation}
Hence
\[
\Lambda_{w(t)}^{-1}B\Lambda_{w(t)}
=e^{t\Lambda}\Lambda_w^{-1}B\Lambda_w e^{-t\Lambda}.
\]
Multiplication on the left by an upper-triangular matrix with
positive diagonal does not change the normalised RQ orthogonal
factor. Consequently
\[
\begin{aligned}
\Phi_t(\Psi)\blacktriangleleft B
&=\operatorname{rq}\bigl(\Psi e^{-t\Lambda}
\Lambda_{w(t)}^{-1}B\Lambda_{w(t)}\bigr)\\
&=\operatorname{rq}\bigl(\Psi\Lambda_w^{-1}B
\Lambda_w e^{-t\Lambda}\bigr)\\
&=\Phi_t(\Psi\blacktriangleleft B).
\end{aligned}
\]
\end{proof}
\end{rem}
\section[Reduction to stochastic group orbits]{Reduction of the Toda flows to the $ST_d(\R)$ orbits}
\label{sect:sect}
We are now going to review the effects of stochastic group action on symmetric matrices: suppose $L=\Psi\Lambda\Psi^T$ and $B\in ST_d(\R)$. Then we are going to describe what canonical form for the matrix $L$ one can get by passing to $L_B=(\Psi\blacktriangleleft B)\Lambda (\Psi\blacktriangleleft B)^T$: it follows from the discussion of the previous section that this action will preserve the trajectories of Toda system, that is, if $L=L(t)$ is a solution of the Toda equation \eqref{eq:Toda1}, then $L(t)_B$ will also satisfy the equation 
\[
\dot L_B=[L_B,M(L_B)].
\]
\subsection{Lanczos map and stochastic group action}
Let us first recall the Lanczos (or Lanczos-Krylov) algorithm for reducing a symmetric matrix to the Jacobi form, the three-diagonal symmetric matrix. Let $L$ be a symmetric $d\times d$ matrix; we can associate with it a self-adjoint linear operator in the Euclidean space $\R^d$ with respect to the standard scalar product. Let us also suppose that $L$ is in general position in the sense that its minimal annulling polynomial is of degree $d$ (and hence coincides with the characteristic polynomial). In this case for a cyclic vector $v\in \R^d$ the basis obtained by the Gram--Schmidt orthogonalisation of the system $v, L(v),\ldots L^{d-1}(v)$ (the so-called \textit{Krylov basis}) provides the Jacobi form for the operator $L$. We introduce the notation for both bases used in this process: let $v_i=L^{i-1}(v)$ and let $u_1,\dots,u_d$ denote the result of the Gram-Schmidt orthogonalisation of $v_1,\dots,v_d$ (in that order). Let us also introduce matrices $V$ and $U$ whose columns are vectors of these bases, written with respect to the standard Euclidean basis in $\R^d$, where we count the vectors from right to left, i.e.:
\bea
V=(v_d,\ldots,v_1)\nn
\eea
and similarly for $U$. Then
\bea
\label{VUR}
V=UR,
\eea
where $R$ is a lower-triangular matrix corresponding to the orthogonalisation procedure and $U$ is the transition matrix from the standard basis to the new orthonormal Krylov basis; in particular, $U$ is an orthogonal matrix itself. We claim that the matrix
\bea
L_J=U^T L U.\nn
\eea 
is a Jacobi (three-diagonal symmetric) matrix: indeed, every vector $u_i$ is equal to a linear combination of $v_i$ and vectors with smaller indices. As $L(v_i)=v_{i+1}$ for $i=1,\dots,d-1$, we see that $L(u_i)=c_i u_{i+1}+\dots$, where $\dots$ denotes a linear combination of $u_1,\dots,u_i$. Hence the matrix of $L$ in the \emph{forward} orthonormal basis $(u_1,\dots,u_d)$ has the form
\[
L_J=\begin{pmatrix}
* & * & * &\dots & *\\
c_1 & * & * &\dots & *\\
0 & c_2 & * &         & *\\
\vdots &    &\ddots &        &\vdots\\
0 & \dots & \dots & c_{d-1}&*
\end{pmatrix}.
\]
The conclusion follows since the matrix is symmetric, hence tridiagonal; reversing the basis keeps it tridiagonal. Below we will use the notation
\bea
L_J=\mathcal{L}_v (L),\nn
\eea
where $v$ denotes the ``initial'' vector in the basis $v,L(v),\dots,L^{d-1}(v)$, which determines the procedure.

Now turn back to the $ST_d(\R)$ group action $\blacktriangleleft$ on
$SO_d(\R)$. Conjugating the diagonal matrix of eigenvalues by an
orthogonal matrix, we transfer this action to the fixed simple-spectrum
isospectral manifold: for $L=\Psi\Lambda\Psi^T$ put
\[
L_B=(\Psi\blacktriangleleft B)\Lambda
(\Psi\blacktriangleleft B)^T.
\]
The formula is independent of the allowed changes of signs of the
ordered eigenvectors $\Psi$, because diagonal sign matrices commute
with $\Lambda$ and the normalised RQ decomposition behaves
compatibly with multiplication by such matrices from the right.

We can now formulate the second main theorem of our paper.
\begin{theorem}\label{Lan}
Let $L=\Psi\Lambda\Psi^T$ be a real symmetric matrix with
simple spectrum and suppose that $e_d$ is a cyclic vector, i.e.
$\det(e_d,Le_d,\ldots,L^{d-1}e_d)\ne0$.
Use the standard Lanczos orthogonalisation starting with $e_d$,
then reverse the orthonormal basis and adjust the sign of its first
vector so that $U\in SO_d(\R)$ and $Ue_d=e_d$.
Then the Jacobi matrix
\[
L_J=\mathcal L_{e_d}(L)=U^TLU
\]
lies in the same stochastic orbit as $L$ on the fixed isospectral
manifold. The signs of its nonzero off-diagonal coefficients are
fixed by this convention.
\end{theorem}
Observe that this theorem gives a simple numerical method of
passing from a generic full symmetric matrix to a tridiagonal
representative of its stochastic orbit.
\begin{proof}
As we have explained earlier, the matrix $U^TLU$ is
symmetric tridiagonal, and the cyclicity condition guarantees that
its subdiagonal entries do not vanish. Since the reversed Lanczos
basis has the last vector equal to $e_d$, and the orientation
correction changes only the first vector, we have $Ue_d=e_d$.
Let $\Psi_J=U^T\Psi$. We see immediately that
\[
\rho(\Psi_J)=e_d^TU^T\Psi=e_d^T\Psi=\rho(\Psi).
\]
By Corollary~\ref{cor:cor}, these two orthogonal matrices
belong to the same stochastic orbit, as we claimed.
\end{proof}
\begin{rem}\label{rem:rem1}
Let us again observe that the direct consequence of the equality
$\rho(\Psi)=\rho(\Psi_J)$ is the following chain of conclusions:
\begin{enumerate}
\item the matrices $\Psi$ and $\Psi_J$ lie in the \textbf{same
fibre} over $\rho(\Psi)\in S^{d-1}_{\times}$;
\item these fibres coincide with the orbits of the stochastic group;
\item consequently the Jacobi matrix
$L_J=\Psi_J\Lambda\Psi_J^T$ is obtained from
$L=\Psi\Lambda\Psi^T$ by the action of a stochastic matrix.
\item Moreover, if $B$ is the stochastic matrix that induces the Lanczos map $L\to L_J$, then as we saw earlier, the map $L(t)\to L(t)_B$ (in notations of the section \ref{sect:sect}) sends the trajectory of the full symmetric Toda system through $L$ to the trajectory of similar system through $L_J$, which is just the trajectory of the open Toda chain, since $L_J$ is Jacobian.
\end{enumerate}
\end{rem}
It follows from our result that the Lanczos procedure is not merely a numerical algorithm:
it selects a particular Jacobi representative of each generic
stochastic orbit. This orbitwise selection should not however be confused with the action of a single group element independent of $L$: obviously the stochastic matrix in Lanczos map depends on the matrix $L$.
\begin{rem}
It is possible to describe explicitly the stochastic matrix $B$
such that $\Psi\blacktriangleleft B=\Psi_J$.
\begin{lemma}\label{B}
The element
\[
B=\Lambda_{\rho(\Psi)}\Psi^T\Psi_J
\Lambda_{\rho(\Psi)}^{-1}
\]
lies in $ST_d(\R)$ and satisfies
$\Psi\blacktriangleleft B=\Psi_J$.
\end{lemma}
\begin{proof}
Put $w=\rho(\Psi)=\rho(\Psi_J)$. Then
\[
\mathbf1^TB=w\Psi^T\Psi_J\Lambda_w^{-1}
=e_d^T\Psi_J\Lambda_w^{-1}=\mathbf1^T,
\qquad \det B=1.
\]
Also
$\Psi\Lambda_w^{-1}B\Lambda_w
=\Psi\Psi^T\Psi_J=\Psi_J$, already orthogonal, and its
normalised RQ-factor is $\Psi_J$.
\end{proof}
\end{rem}
\begin{rem}
Observe that when the eigenvalues of $L$ are distinct and 
the bottom row $w_1,\dots,w_d$ of the diagonalizing matrix $\Psi$ is nonzero and fixed, the Jacobi matrix in stochastic orbit of $L$ is unique \textit{after fixing the signs} of its nonzero off-diagonal entries. In effect, the spectral moments $\sum_iw_i^2\lambda_i^k$ determine the standard Jacobi recurrence,
and diagonal sign conjugations fixing $e_d$ produce the other signed representatives. The convention in Theorem~\ref{Lan} selects one such representative on each connected component of
the cyclic locus; see \cite{Moser1975}.
\end{rem}
\begin{rem} \label{rem:rem2} In effect one can show that Lanczos map intertwines the trajectories of Toda system: it follows from the uniqueness of this representation and from the observation that the action of stochastic group preserves the trajectories (see remark \ref{rem:rem1}, part 4). But one can also prove it directly:
\begin{prop}\label{prop:intertwines}
Let $L(t)$ solve the full symmetric Toda equation with $e_d$ cyclic at $t=0$, and let $L_J(t)$ be the open Toda solution with $L_J(0)=\mathcal L_{e_d}(L(0))$. For the sign convention of Theorem~\ref{Lan},
\begin{equation}\label{eq:Lanczosintertwining}
\mathcal L_{e_d}(L(t))=L_J(t)
\end{equation}
for every finite $t$ for which these solutions are defined.
\end{prop}
\begin{proof}
By \eqref{eq:wtformula}, the squares of coefficients 
$w_i(t)^2$ of the the bottom row of the matrix $\Psi(t)$ (solution of the Toda equation on $SO_d(\R)$) 
are
\[
w_i(t)^2=\frac{w_i(0)^2e^{-2\lambda_i t}}
{\sum_jw_j(0)^2e^{-2\lambda_jt}}.
\]
This is the same evolution for the full Toda solution and the tridiagonal open Toda solution, since both satisfy the same isospectral Lax equation. Tridiagonal matrices form an invariant submanifold and the nonzero off-diagonal entries preserve their signs under the flow. By the finite inverse spectral correspondence for Jacobi matrices with prescribed off-diagonal signs \cite{Moser1975}, the two tridiagonal representatives coincide.\end{proof}
\end{rem}

\subsection{Solution for the full symmetric system}
The results of previous sections provide a natural way to reduce
the solution of the full symmetric Toda system to the solution of
the open Toda chain. Due to Theorem~\ref{Lan},
$\mathcal L_{e_d}(L)$ is a Jacobi matrix; moreover, by remark \ref{rem:rem1} part 4 and 
remark \ref{rem:rem2} Proposition~\ref{prop:intertwines}, solving the open chain through
this initial Jacobi representative gives the Lanczos reduction of
the full trajectory. Let us formulate this as a separate statement:
\begin{theorem}\label{thm:restore}
Let $L_0=\Psi_0\Lambda\Psi_0^T$ have simple spectrum and cyclic
vector $e_d$, and put
$L_J(0)=\mathcal L_{e_d}(L_0)$.
Let $B_0$ be the stochastic element of Lemma~\ref{B} at $t=0$.
If $\Psi_J(t)$ is the lifted solution through
$\Psi_J(0)=\Psi_0\blacktriangleleft B_0$,
then
\[
\Psi(t)=\Psi_J(t)\blacktriangleleft B_0^{-1},
\qquad L(t)=\Psi(t)\Lambda\Psi(t)^T
\]
is the full symmetric Toda solution with initial condition $L_0$.
\end{theorem}
\begin{proof}
Apply the equivariance of Proposition~\ref{prop:equivariance}
to the fixed group element $B_0^{-1}$. At $t=0$ the
result is $\Psi_0$. Uniqueness for the lifted Toda equation
proves the assertion.
\end{proof}
Let us conclude this part of the paper by an important observation.
The reconstruction of the full Toda system trajectory from the open Toda chain one that we've just described (i.e. with the help of the inverse stochastic matrix action), uses a matrix of eigenvectors
and hence entails an algebraic spectral computation. However,
there is another approach which avoids the extraction of the
individual eigenvalues and eigenvectors after the open Toda solution
has been found, albeit on the expense of solving a series of equations separately for every value of time $t$.

Namely, suppose that $J_0=\mathcal L_{e_d}(L_0)=U_0^TL_0U_0$ and
that $J(t)$ is the corresponding open Toda solution. Write
\[
J_0=\begin{pmatrix}
a_1&b_1&&\\b_1&a_2&\ddots&\\&\ddots&\ddots&b_{d-1}\\
&&b_{d-1}&a_d
\end{pmatrix},\qquad
J(t)=\begin{pmatrix}
A_1&B_1&&\\B_1&A_2&\ddots&\\&\ddots&\ddots&B_{d-1}\\
&&B_{d-1}&A_d
\end{pmatrix},
\]
where the nonzero $b_i$ and $B_i$ have identical signs.
We define a sequence of column vectors $r_k(t)$ by the
three-term recursion
\begin{equation}\label{eq:intertwiningrecursion}
\begin{aligned}
r_1(t)&=e_1,\\
r_2(t)&=b_1^{-1}(J(t)-a_1I)r_1(t),\\
r_{k+1}(t)&=b_k^{-1}\bigl((J(t)-a_kI)r_k(t)-b_{k-1}r_{k-1}(t)\bigr),
\quad 2\le k\le d-1.
\end{aligned}
\end{equation}
\begin{lemma}\label{lem:intertwiner}
The matrix $R(t)=(r_1(t),\ldots,r_d(t))$ is upper triangular,
its diagonal entries are positive, and
\begin{equation}\label{eq:intertwineR}
J(t)R(t)=R(t)J_0,\qquad
R_{kk}(t)=\prod_{j=1}^{k-1}\frac{B_j(t)}{b_j}>0.
\end{equation}
\end{lemma}
\begin{proof}
The first $d-1$ column equations are precisely
\eqref{eq:intertwiningrecursion}. The tridiagonality of $J(t)$
shows inductively that $r_k$ is supported in its first $k$
coordinates; its $k$th coordinate is the displayed product.
For the final column equation, denote by $p_k(x)$ the monic
characteristic polynomial of the leading $k\times k$ principal
block of $J_0$. The usual three-term recurrence gives
\[
r_k=\frac{p_{k-1}(J(t))e_1}{b_1\cdots b_{k-1}},
\qquad
p_k(x)=(x-a_k)p_{k-1}(x)-b_{k-1}^2p_{k-2}(x).
\]
The last column equation is equivalent to $p_d(J(t))e_1=0$,
which follows from Cayley--Hamilton, since $J(t)$ is isospectral
to $J_0$.
\end{proof}
\begin{theorem}\label{thm:lifting}
Let $L_0$ have cyclic vector $e_d$ and let
$J_0=U_0^TL_0U_0$ be its reversed Lanczos representative,
with $U_0\in SO_d(\R)$ and $U_0e_d=e_d$.
Let $J(t)$ solve the open Toda equation with $J(0)=J_0$.
Build $R(t)$ using \eqref{eq:intertwiningrecursion}, and
form the upper-triangular Cholesky factor $S(t)$ with positive
diagonal:
\begin{equation}\label{eq:Cholfactor}
S(t)^TS(t)=U_0R(t)^TR(t)U_0^T.
\end{equation}
Then the full Toda solution with initial condition $L_0$ is
\begin{equation}\label{eq:finalformula}
L(t)=S(t)L_0S(t)^{-1}.
\end{equation}
Equivalently, if $R(t)U_0^T=Q(t)S(t)$ is its QR-decomposition,
\[
L(t)=Q(t)^TJ(t)Q(t).
\]
\end{theorem}
\begin{proof}
The intertwining equation gives
$J(t)=R(t)J_0R(t)^{-1}$. Applying the QR factorisation
of $R(t)U_0^T$ proves the second expression for $L(t)$.
To identify the first expression with the Toda trajectory,
use the classical QR solution \cite{S1980,TCS23}:
\[
e^{tJ_0}=Q_J(t)T_J(t),\qquad
J(t)=T_J(t)J_0T_J(t)^{-1},
\]
where $T_J(t)$ is upper triangular with positive diagonal.
Both $R(t)$ and $T_J(t)$ satisfy the same intertwining
relation. Such an upper-triangular matrix is uniquely determined
by its first column (use the three-term recurrence), so
$R(t)=T_J(t)/(T_J(t))_{11}$. If
$c(t)=(T_J(t))_{11}>0$, then
\[
U_0R(t)^TR(t)U_0^T=c(t)^{-2}e^{2tL_0}.
\]
Write $e^{tL_0}=Q_L(t)T_L(t)$ for the positive-diagonal QR
decomposition. Then $T_L(t)^TT_L(t)=e^{2tL_0}$, and uniqueness
of Cholesky factorisation implies $S(t)=c(t)^{-1}T_L(t)$.
Therefore \eqref{eq:finalformula} is the standard QR formula
$L(t)=T_L(t)L_0T_L(t)^{-1}$ for the full symmetric Toda
system. At $t=0$ one has $R(0)=I$ and $S(0)=I$.
\end{proof}
\begin{rem}
The intertwining matrix $R(t)$ is obtained by polynomial
expressions in the entries of $J(t)$, with coefficients rational
in the fixed nonzero subdiagonal entries of $J_0$.
The QR and Cholesky steps involve square roots, so the lifting
is algebraic, not strictly polynomial. This concerns the
\emph{reconstruction step} once $J(t)$ has been found:
solving the open chain itself may still rely on spectral methods.
\end{rem}

\section{Concluding remarks and further questions}
We have described a stochastic group action on an open part
of $SO_d(\R)$, identified its orbits with the fibres of the
projection to the last row, and interpreted the Lanczos algorithm
as a choice of Jacobi representative on each generic orbit.
The group action is compatible with the full symmetric Toda flow, and on the infinitesimal level coincides with the Lie algebra action that appeared in prior work on Lie--Bianchi integrability.
The inverse passage from the open chain can be made without
individual eigenvalue extraction. 

Several natural questions remain. How does the RQ-adapted action
compare globally with the Lie--Bianchi realisations of \cite{CST26}?
How does the geometry of the stochastic orbit space extend across
the singular strata where $e_d$ ceases to be cyclic? Can these
constructions be adapted to other root systems or to other
Hessenberg-type reductions? What other structures that exist on open Toda chain systems (like the higher compatible Poisson structures or Henkel determinant solution formulas) can be transferred to the full Toda system with the help of this method? Finally, is there an operator-theoretic
or quantum analogue of the Lie--Bianchi symmetry criterion?

\medskip
\paragraph{Acknowledgements}{The work of Yu.~Ch., (who contributed to sections 3.1 and 3.3) was supported by the state assignment of the NRC ``Kurchatov institute''; the work of G.~Sh. (sections 3.2 and 4.1) was supported by the Russian Science Foundation under grant no. 25-11-00-210; the work of D.~T. (sections 3.4 and 4.2) was carried out within the framework of a development program for the Regional Scientific and Educational Center of the P.~G.~Demidov Yaroslavl State University with financial support from the Ministry of Science and Higher Education of the Russian Federation (Agreement on provision of subsidy from the federal budget No. 075-02-2026-1331) and was partially supported by a grant Leader Mathematics \#25-7-1-4-1 of the Basis foundation.}

\section*{Declaration of generative AI use}
OpenAI's ChatGPT was used for exploratory discussions of
mathematical arguments, assistance with symbolic computations,
and preparation of the \LaTeX{} manuscript.
The authors independently take responsibility for all mathematical
statements, proofs, references and the final text.

\end{document}